\documentclass[aps,prl,twocolumn,superscriptaddress]{revtex4-1}
\usepackage{amsmath}
\usepackage{graphicx}
\usepackage{esint}
\usepackage{epstopdf}%This line makes .eps figures into .pdf - please comment out if not required.
\graphicspath{{pict/}{./}}
\usepackage{bm}%
\newcounter{Fig}

\newcommand{\be}{\begin{equation}}
\newcommand{\ee}{\end{equation}}

\begin{document}

%\title{Q-factor enhancement for all-dielectric nanoresonators through relaxed total internal reflection}
\title{Cascaded Rotational Doppler Effect}
\author{Junhong Deng}
\author{King Fai Li}
\affiliation{Department of Materials Science and Engineering, Southern University of Science and Technology,
Shenzhen, P. R. China}
\author{Wei Liu}
\email{wei.liu.pku@gmail.com}
\affiliation{College for Advanced Interdisciplinary Studies, National University of Defense Technology, Changsha, P. R. China}
\author{Guixin Li}
\email{ligx@sustc.edu.cn}
\affiliation{Department of Materials Science and Engineering, Southern University of Science and Technology,
Shenzhen, P. R. China}
\affiliation{Shenzhen Institute for Quantum Science and Engineering, Southern University of Science and Technology, Shenzhen, P. R. China}

\begin{abstract}
We propose and substantiate experimentally the cascaded rotational Doppler effect for interactions of spinning objects with light carrying angular momentum.  Based on the law of parity conservation for electromagnetic interactions, we reveal that the frequency shift can be doubled through cascading two rotational Doppler processes which are mirror-imaged to each other. This effect is further experimentally verified with a rotating half-wave plate, and the mirror-imaging process is achieved by reflecting the frequency-shifted circularly polarized wave upon a mirror with a quarter-wave plate in front of it. The mirror symmetry and thus parity conservation guarantees that this doubled frequency shift can be further multiplied with more successive mirror-imaging conjugations, with photons carrying spin and/or orbital angular momentum, which could be widely applied for detection of rotating systems ranging from molecules to celestial bodies with high precision and sensitivity.
\end{abstract}
\maketitle

Similar to its translational counterpart, the rotational Doppler effect (RDE) originates from the angular momentum and energy exchange between rotating objects and waves carrying angular momentum (of either spin or orbital forms) that interact with each other~\cite{ALLEN_2003__Opticala,GARETZ_1979_Opt.Commun._Variable,GARETZ_1981_JOSA_Angular,SIMON_1988_Phys.Rev.Lett._Evolving,NIENHUIS_1996_Opt.Commun._Doppler,
COURTIAL_1998_Phys.Rev.Lett._Measurementa,COURTIAL_1998_Phys.Rev.Lett._Rotational}. Since its preliminary experimental verifications~\cite{ALLEN_2003__Opticala}, the RDE has been verified to be a quite generic phenomenon~\cite{ALLEN_1994_OpticsCommunications_Azimuthal,BIALYNICKI-BIRULA_1997_Phys.Rev.Lett._Rotationala}.  Such an effect can be employed for rotating frequency detections for objects at various scales spanning from microscopic molecules to macroscopic objects~\cite{MICHALSKI_2005_Phys.Rev.Lett._Experimentala,KORECH_2013_Nat.Photonics_Observing,LAVERY_2013_Science_Detectiona}, which can be potentially further applied for cosmoscopic observations and measurements in astronomy.

One of the most severe obstacles for further applications of RDE is that the accompanying rotational Doppler shift (RDS) is generally too tiny to be detected with a high precision, especially for spinning objects with a relatively low rotating frequency~\cite{ALLEN_2003__Opticala}. Conventional methods that can partially obstacle this problem rely on twisted photons with higher orbital angular momentum or a nonlinear process with higher harmonic generations, through which larger angular momentum transfers and thus enhanced RDS can be obtained~\cite{ALLEN_2003__Opticala,LAVERY_2013_Science_Detectiona,SIMON_1968_Phys.Rev._Secondharmonic,LI_2016_Nat.Phys._Rotational,FAUCHER_2016_Phys.Rev.A_Rotational,GEORGI_2017_OpticaOPTICA_Rotational,LI_2018_LaserPhotonicsRev._Observation,XIAO_2018_Opt.ExpressOE_Orbital}. Nevertheless,  those approaches have their own drawbacks, including the requirement of complicated optical elements for sophisticated wave shaping,  significant intensity dependence, and low conversion efficiencies, to name but a few. At the same time, it is easy to identify a common feature and limitation of almost all conventional approaches: only a single-round energy and angular momentum exchange has been exploited, that is, photons extract energy from or lose energy to the spinning bodies only once. It is natural to expect a manyfold enhancement of the RDS, if more consecutive rounds of energy exchange can be synchronized in an incremental way.

In this work we propose the concept of cascaded RDE through conjugating the conventional rotational Doppler process with its mirror-imaging counterpart. The spatial inversion invariance and thus law of parity conservation~\cite{LEE_1956_Phys.Rev._Question,BARRON_2009__Molecular,SHANKAR_2011__Principles} secures in this cascaded process a twofold enhancement of the RDS. This doubled RDS  at the same time can be further multiplied accordingly, through more recursive mirror imaging conjugations. We further experimentally demonstrate this twofold RDS enhancement by interacting circularly polarized (CP) light with a rotating half-wave plate (HWP). The original rotational Doppler process is cascaded over its subsequent mirror-imaging counterpart, which is realized by reflecting back the frequency-shifted beam upon a mirror with an extra quarter-wave plate (QWP) in front of it. we expect the concept that has been proposed and substantiated can be widely employed to improve the sensitivity and accuracy of RDE based sensing and measurements, which can pervade various fields involving dynamic rotating systems.

\begin{figure}[tp]
\centerline{\includegraphics[width=8cm]{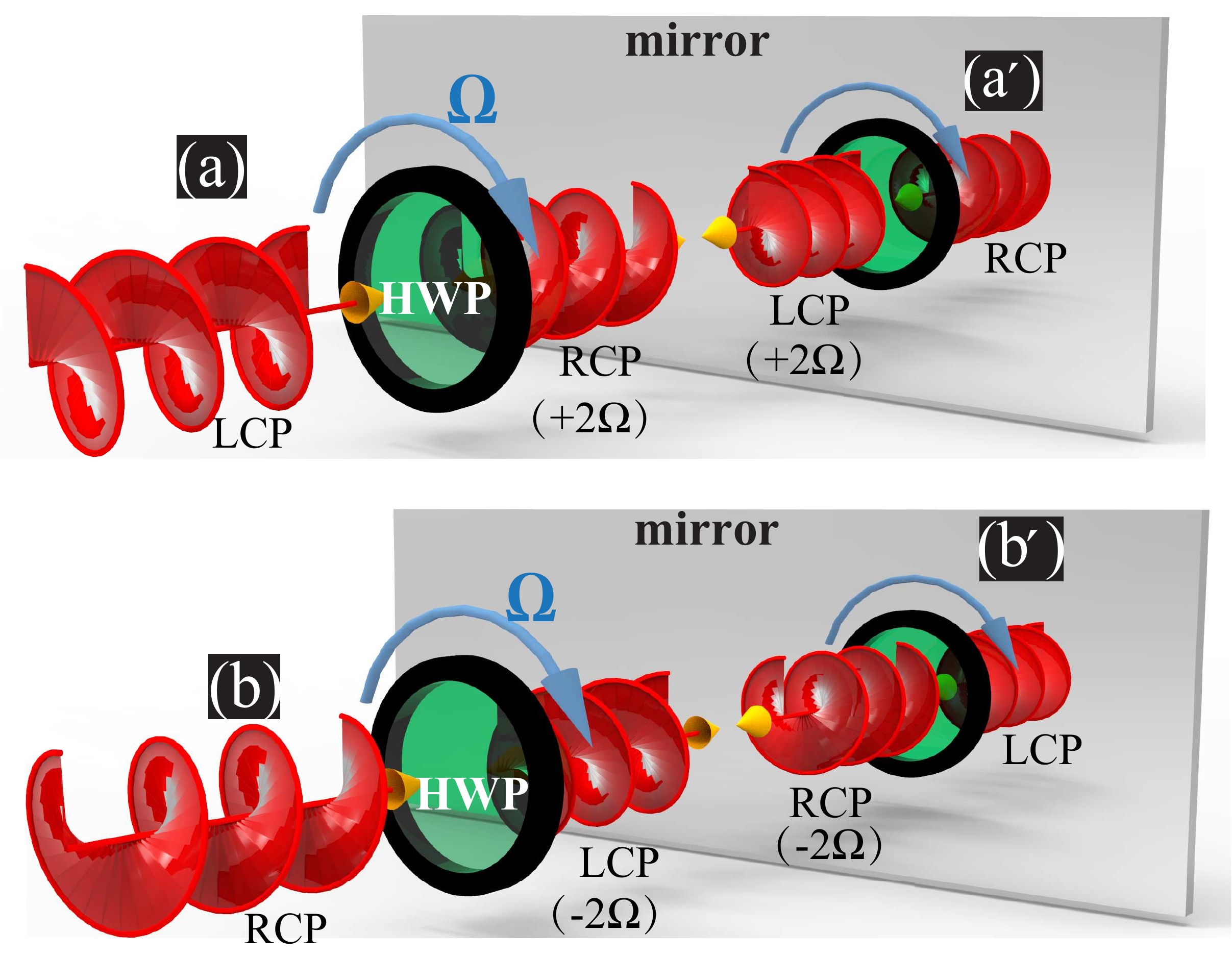}} \caption {\small Original rotational Doppler processes with CP light are shown in (a) and (b), with the coresponding mirror-imaging counterparts shown in (a$^{\prime}$) and (b$^{\prime}$), respectively.  The angular momentum  (pseudo-vector) is maintained (for normal incidence) withstanding the mirror-imaging operation, while both the mentum (wave vector) and handedness of the circularly polarized light are reversed. During both processes that are mirror-imaged to each other, the parity conservation law guarantees that identical RDS  will be experienced, with HWPs rotating at angular frequency $\Omega$.}\label{fig1}
\end{figure}

The conventional single-round RDE with CP light is shown schematically in Figs.~\ref{fig1} (a) and (b), where we adopt the convention that the angular momentum of left/right-handed circularly polarized (LCP/RCP) light is anti-parallel/parallel to the propagation direction $\textbf{k}$. When transmitted through a rotating half-wave plate (HWP) with angular velocity $\Omega=2\pi f$ that is anti-parallel to $\textbf{k}$, besides the handedness flipping [see Figs~\ref{fig1}(a) and(b)], the incident LCP and RCP waves will experience respectively the RDS of $\pm2\Omega$~\cite{GARETZ_1979_Opt.Commun._Variable,GARETZ_1981_JOSA_Angular}. The mirror-imaging counterparts of the processes in Figs.~\ref{fig1} (a) and (b) are shown in Figs.~\ref{fig1} (a$^{\prime}$) and (b$^{\prime}$), respectively. Though the mirror-imaging operation will reverse the handedness of both incident and rotational Doppler shifted waves, according to the law of parity conservation~\cite{LEE_1956_Phys.Rev._Question,BARRON_2009__Molecular,SHANKAR_2011__Principles}, during the original and mirror imaging processes identical RDS of either $+2\Omega$ or $-2\Omega$ should be experienced, as is shown in Fig.~\ref{fig1}. The same conclusion could be drawn by more detailed analysis based on  conservations of angular momentum and energy ~\cite{GARETZ_1979_Opt.Commun._Variable,GARETZ_1981_JOSA_Angular,ALLEN_1994_OpticsCommunications_Azimuthal}, which nevertheless is by far not as direct or intuitive as the left-right symmetry argument provided here (refer to supplemental materials \cite{Supplemental_Material} for more details). To obtain the RDS doubling, it is obvious from Fig.~\ref{fig1} that we can cascade two mirror-imaging processes [Fig.~\ref{fig1}(a$^{\prime}$) subsequent to Fig.~\ref{fig1}(a); Fig.~\ref{fig1}(b$^{\prime}$) subsequent to Fig.~\ref{fig1}(b)]. In contrast, if two processes that are not mirror-imaged to each other [Fig.~\ref{fig1}(a$^{\prime}$) subsequent to Fig.~\ref{fig1}(b); Fig.~\ref{fig1}(b$^{\prime}$) subsequent to Fig.~\ref{fig1}(a)] are cascaded, two opposite shifts cancel each other, ending up with null overall RDS.

 %($\lambda_{A-G}/R=6.16,~9.54,~4.58,~3.54,~9.54,~3.54$, and $6.07$)

The most conventional approach for mirror-imaging conjugation can be implemented by a direct mirror reflection operation. For CP light upon reflection (normal incidence) by a conventional mirror with isotropic response, the handedness is reversed, while the angular momentum is unchanged due to its pseudovector nature and the rotational symmetry of the mirror~\cite{SHANKAR_2011__Principles,BIRSS_1964__Symmetry}. As a result, unfortunately, the processes that have been cascaded through reflecting by a conventional mirror are not mirrored imaged to each other [Fig.~\ref{fig1} (a) plus Fig.~\ref{fig1}~(b$^{\prime}$), or Fig.~\ref{fig1} (b) plus Fig.~\ref{fig1}(a$^{\prime}$)] and thus no RDS can be observed. According to Fig.~\ref{fig1}, a mirror-imaging conjugation requires a reflection process that maintains the handedness and thus reverses the angular momentum of incident CP light, which can be realized simply by placing a QWP in front of the conventional mirror. This is due to the fact that a round trip through the QWP would induce an extra $\pi$-phase difference for the two orthogonal linear polarizations, which subsequently reverses the spin angular momentum and preserves the handedness of the reflected light. In a word, the incorporation of an extra QHP enables an effective mirror imaging conjugation and thus the observation of cascaded RDE with doubled RDS.
\begin{figure}[tp]
\centerline{\includegraphics[width=8.9cm]{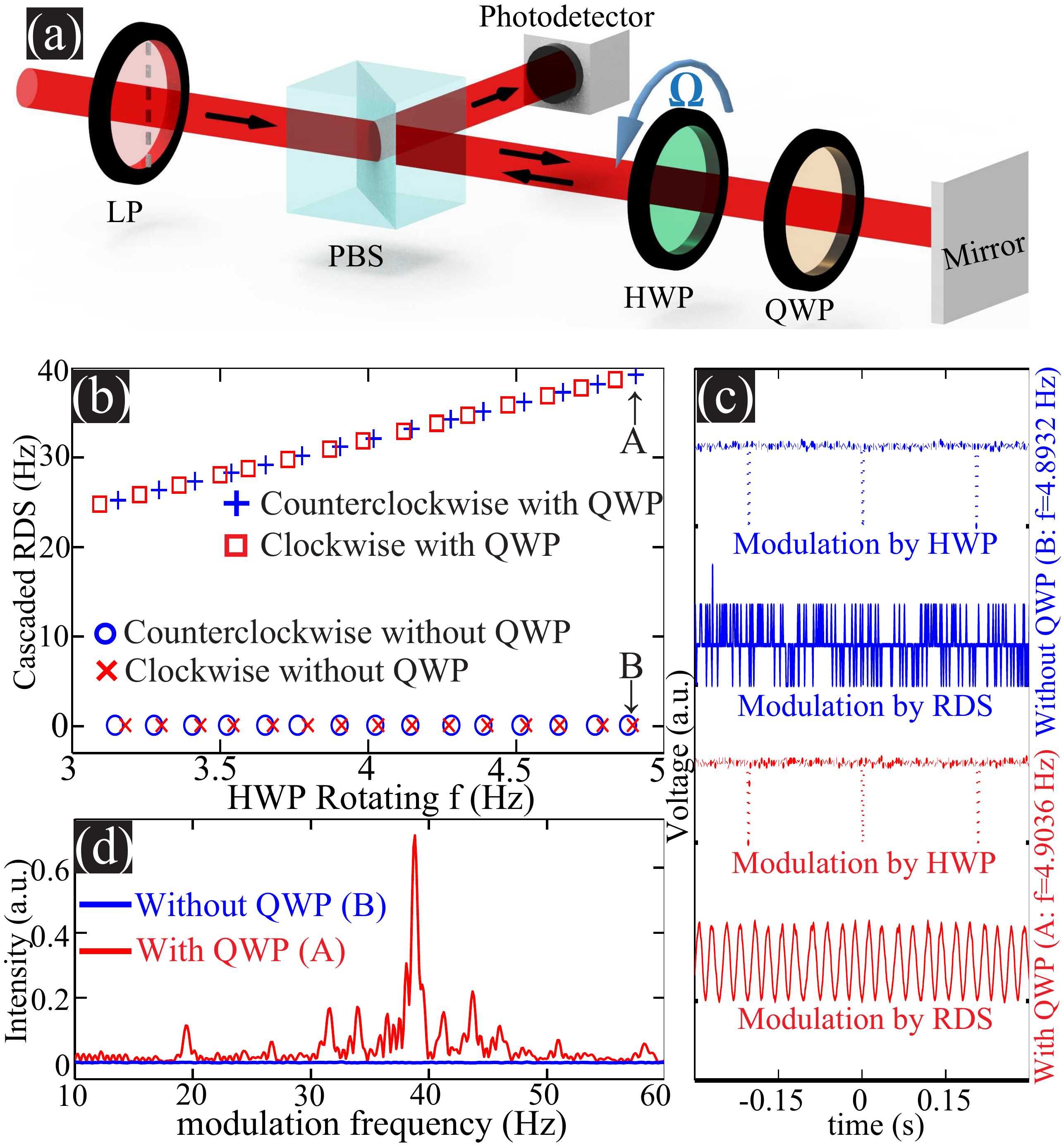}} \caption {\small (a) Experimental setup: the  $632.8$nm He-Ne laser beam is made linearly polarized befored being directed through the PBS, rotating HWP, the QWP and then reflected back by a conventional mirror; the reflected beam is then split to a photodetector after passing through the rotating HWP again. (b) Dependence of the cascaded RDS on the rotating frequency of the HWP (including counterclockwise and clockwise rotations), where both sets of results with and without the QWP are shown. Two representative points (\textbf{A} and \textbf{B} with $f_{\rm {\textbf{A}}}=4.9036$ Hz and $f_{\rm {\textbf{B}}}=4.8932$ Hz) are indicated in (a) and the corresponding temporal and spectral responses (characterized by signal modulations) recorded by the photodetector at those points are show respectively in (c) and (d). In (c) for clarity the signal modulation of frequency $f$ induced solely by the HWP rotating at the same frequency are also presented.}\label{fig2}
\end{figure} %($\lambda_{A-G}/R=6.16,~9.54,~4.58,~3.54,~9.54,~3.54$, and $6.07$)

As a next step, we experimentally verify what has been proposed above with the experimental setup shown in Fig.~\ref{fig2}(a), of which more details can be found in the supplemental materials \cite{Supplemental_Material}. A linearly polarized (LP) beam from a He-Ne laser propagates through a rotating quartz HWP [angular velocity $\Omega$; the rotating direction shown in Fig.~\ref{fig2}(a) is defined as counterclockwise] and then is reflected back by a conventional metallic mirror with a QWP in front of it. After a round-trip through the rotating HWP the beam is directed into a photodetector by a polarized beam splitter (PBS). After the first transmission through the rotating HWP, the incident LP beam is divided into waves of opposite handedness, which experience opposite RDS of $\pm2\Omega$ respectively. According to the analysis based on Fig.~\ref{fig1}, when transmitted through the HWP again, the reflected wave with preserved handedness would experience the same frequency shift, ending up with total RDS of $\pm4\Omega$ due to the mirror-imaging conjugation. This would enable a total beam beating rate of $8\Omega$ that can be directly observed by the photodetector, which is independent of the rotating direction (counterclockwise or clockwise) of the HWP. In contrast, when the QWP is removed, during the cascaded processes that are not mirror-imaged to each other the RDE cancel each other and thus no effective frequency beating effect would be detected.

The experimentally measured cascaded RDS with different HWP rotating frequencies (including both counterclockwise and clockwise rotations) are show in Fig.~\ref{fig2}(b), where both sets of results (with or without the QWP in front of the mirror) are included. For both cases the results (beating frequencies that are twice of the cascaded RDS) agree well with the theoretically predicted shifts, indicating RDS of $\pm4\Omega$ for mirror-imaging cascading and null for non-mirror-imaging cascading.  To be more specific, for each case a representative point is indicated in Fig.~\ref{fig2}(b) (points \textbf{A} and \textbf{B} for both cases with and without QWP, respectively; $f_{\rm {\textbf{A}}}=4.9036$ Hz and $f_{\rm {\textbf{B}}}=4.8932$ Hz). The corresponding temporal beam beating phenomena (characterized by signal modulations) observed with the photodetector is shown in Fig.~\ref{fig2}(c), where for better comparison we have also shown the  pure HWP rotation induced modulations of the signal with a frequency of $f$. Figure~\ref{fig2}(c) conveys the clear message that mirror-imaging conjugation (with QWP) will induce doubled RDS of $4f$ (steady signal modulations with a frequency of $8f$ which is the beating frequency). In contrast, such a RDS would be eliminated with the QWP removed, where the steady signal modulations disappear, as is shown in Fig.~\ref{fig2}(c). The extracted spectral information from the temporal responses shown in Fig.~\ref{fig2}(c) is presented in Fig.~\ref{fig2}(d), which further confirms our conclusions drawn above.

An alternative interpretation of the RDS doubling observed can be obtained through implementing the time inversion operation~\cite{SHANKAR_2011__Principles,POTTON_Rep.Prog.Phys._reciprocity_2004}. A direct examination of the processes shown in Fig.~\ref{fig1} tells that, if any of the rotating direction of the HWP is reserved, the process would be rendered to be time-reversal-symmetric to its originally mirror-imaging counterpart. The time reversal symmetry would eliminate the overall RDS if two time-reversal-symmetric processes are conjugated, since the back-propagating time-reversed wave (also can be induced by the mirror-QWP doublet for the phase-conjugated reflection) would be required to return to its original state with the same frequency~\cite{SHANKAR_2011__Principles,POTTON_Rep.Prog.Phys._reciprocity_2004,CHEW_1985_JOptSocAmA_Phaseconjugate}. Nevertheless, when the rotating direction of the HWP is fixed during the cascaded processes (as is the case in our experiment shown in Fig.~\ref{fig2}), the time reversal symmetry is broken with the RDS accumulated in an additive way, which is similar to the successive Faraday rotations to the same direction with fixed external magnetic field~\cite{RINARD_1971_AmericanJournalofPhysics_Faraday}.

\begin{figure}[tp]
\centerline{\includegraphics[width=8.5cm]{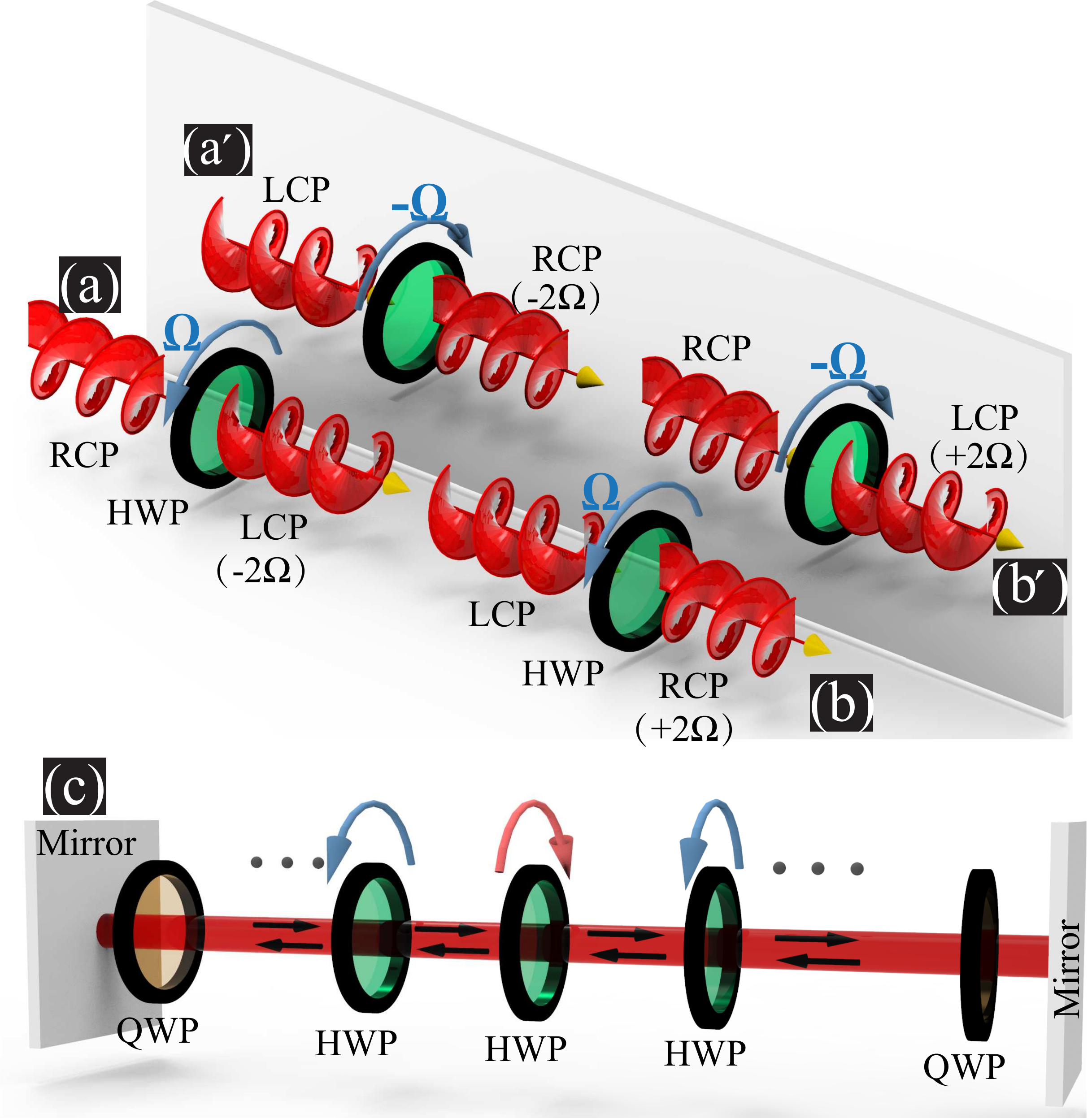}} \caption {\small Original rotational Doppler processes shown in (a) and (b), with the coresponding mirror-imaging counterparts shown in (a$^{\prime}$) and (b$^{\prime}$), respectively. In the two mirror-imaging processes: the wave vector is fixed;  HWPs are rotating to opposite directions;  incident and transmitted CP waves are of opposite handedness and angular momentum. A more sophisticated setup, which can induced many-fold RDS enhancement by synchronising two mirror-imaging-conjugation scenarios of fixed propagation direction or fixed HWP rotating direction, is proposed and shown in (c).  A series of counter-rotating HWPs are placed inside a Fabry-Perot cavity formed by two parallel mirrors, and the handedness preserving reflection processes can be realized either with extra QWPs or through employing anisotropic metamirrors.}\label{fig3}
\end{figure} %($\lambda_{A-G}/R=6.16,~9.54,~4.58,~3.54,~9.54,~3.54$, and $6.07$)

Up to now we have cascaded two mirror-imaging processes that involve two counter-propagating beams, where a reflection operation is required. At the same time, for a beam with fixed propagation direction, we can also achieve the RDS doubling by cascading two mirror-imaging processes shown in Fig.~\ref{fig3} [Fig.~\ref{fig3}(a$^{\prime}$) subsequent to Fig.~\ref{fig3}(a); Fig.~\ref{fig3}(b$^{\prime}$) subsequent to Fig.~\ref{fig3}(b)]. The trade-off to get rid of the reflection mirror is that two parallel counter-rotating HWPs are required for RDS doubling, as then the beam would extract energy from the two plates successively in an additive way~\cite{LI_2016_Nat.Phys._Rotational}. Moreover, the two scenarios [fixed propagation direction shown in Figs.~\ref{fig3}(a)-(b$^{\prime}$) or fixed HWP rotating direction shown in Fig.~\ref{fig1}] for mirror-imaging-conjugation can be further synchronized in a setup shown schematically in Fig.~\ref{fig3}(c): a series of parallel rotating HWPs (rotating directions are opposite for neighbouring HWPs) are placed inside a Fabry-Perot cavity formed by two  mirror-QWP doublets; during each reflection process, the handedness of the CP wave should be preserved, which can be also realized with simpler anisotropic metamirrors~\cite{XIAO_2016_Adv.Opt.Mater._Helicity,jahani_alldielectric_2016,KUZNETSOV_Science_optically_2016,LIU_Phys.Rev.Lett._generalized_2017-1}. Within such a proposed configuration, each time the CP wave passes through the rotating QWP, the RDS would be accumulated  in an additive way. After many rounds of reflections within such a cavity,  manyfold enhancement for the RDS is expected, which in principle is only constrained by the quality factor of the  Fabry-Perot cavity, and the effective functioning spectral bandwidths of the HWPs and QWPs.

In conclusion, here we have proposed and demonstrated a cascaded rotational Doppler effect by successively conjugating  mirror-imaging rotational Doppler processes. For such an effect, the law of parity conservation guarantees that each time the CP light passes through the rotating HWP,  the energy will be exchanged in an accumulative manner, leading to manyfold enhancement for the overall RDS. Here in this study, we have confined other discussions to CP light carry spin angular momentum and rotating HWPs. Similar investigations can be certainly extended to structured light carrying orbital (or both spin and orbital) angular momentum ~\cite{ALLEN_2003__Opticala} passing through more sophisticated rotating bodies, considering the dynamic geometric phase nature of such an effect~\cite{SIMON_1988_Phys.Rev.Lett._Evolving,BLIOKH_2006_Phys.Rev.Lett._Geometrical}. It is expected that the cascaded rotational Doppler effect we have proposed here, together with its translational counterpart, can potentially incubate various highly precise approaches that can be employed for sensing and detection applications, not only for electromagnetic waves, but also for waves of other forms~\cite{GIBSON_2018_Proc.Natl.Acad.Sci._Reversal,LONG_2018_PNAS_Intrinsic}. Moreover, the fusion of cascaded RDE into the recently flourishing fields of time-modulation induced non-reciprocal photonics~\cite{SOUNAS_2017_Nat.Photonics_Nonreciprocal} and non-Hermitian optics~\cite{FENG_Nat.Photonics_nonhermitian_2017} can potentially provide extra flexibilities for the designs of ultra-compact and energy efficient optical devices and circuits.

This research was supported by the National Natural Science Foundation of China (Grant No. 11774145 and 11404403), Guangdong Provincial Innovation and Entrepreneurship Project (Grant 2017ZT07C071), Applied Science and Technology Project of Guangdong Science and Technology Department (2017B090918001) and Natural Science Foundation of Shenzhen Innovation Committee (JCYJ20170412153113701).

%\section*{References}
%\bibliographystyle{osajnl}
%\bibliography{References_scattering}
%merlin.mbs apsrev4-1.bst 2010-07-25 4.21a (PWD, AO, DPC) hacked
%Control: key (0)
%Control: author (8) initials jnrlst
%Control: editor formatted (1) identically to author
%Control: production of article title (-1) disabled
%Control: page (0) single
%Control: year (1) truncated
%Control: production of eprint (0) enabled
%

%==========================
\end{document}